\documentclass[twocolumn,showpacs,preprintnumbers,amsmath,amssymb]{revtex4}

\usepackage{graphicx}
\usepackage{dcolumn}
\usepackage{bm}

\begin{document}

\preprint{APS/123-QED}

\title{ Spin density wave and superconductivity in CaFe$_{1-x}$Co$_{x}$AsF studied by nuclear magnetic resonance }

 \author{ S. Tsutsumi$^1$, N. Fujiwara $^1$ \footnote { Corresponding author: naoki@fujiwara.h.kyoto-u.ac.jp},
S. Matsuishi$^{2, 3}$, and H. Hosono$^{2, 3}$
 }

\affiliation{$^1$ Graduate School of Human and Environmental
Studies, Kyoto University, Yoshida-Nihonmatsu-cyo, Sakyo-ku, Kyoto
606-8501, Japan}


\affiliation {$^2$ Materials and Structures Laboratory (MSL), Tokyo
Institute of Technology, 4259 Nagatsuda,  Midori-ku, Yokohama
226-8503, Japan \ \\ $^3$ Frontier Research Center (FRC), Tokyo
Institute of Technology, 4259 Nagatsuda, Midori-ku, Yokohama
226-8503, Japan}




\date{November 30 2010}


\begin{abstract}

 We performed nuclear magnetic resonance measurements to investigate the evolution of spin-density-wave
 (SDW)  and  superconducting (SC) states upon electron doping  in
 CaFe$_{1-x}$Co$_{x}$AsF, which exhibits an intermediate phase diagram between those of LaFeAsO$_{1-x}$F$_x$ and
 Ba(Fe$_{1-x}$Co$_x$)$_2$As$_2$. We found that homogeneous coexistence of the incommensurate SDW and SC states occurs only in a narrow
 doping region around the crossover regime, which supports $S_{+-}$-wave symmetry. However, only the structural phase transition survives upon further doping, which agrees with predictions from orbital fluctuation theory. The transitional features upon electron doping imply that both spin and orbital fluctuations are involved in the superconducting mechanism.

\end{abstract}

\maketitle

In strongly correlated electron systems, electronic states in a crossover regime between antiferromagnetic (AF) and superconducting (SC) phases have attracted significant research interest. The problem has been often discussed in iron-based superconductors in which the AF and SC states can coexist homogeneously $^{1, 2}$. In iron-based superconductors, most of the parent compounds are in AF phases at low temperatures. A representative AF spin configuration is a stripe-type spin-density-wave (SDW) state arising from nesting between electron and hole pockets $^3$. In a crossover regime, several theoretical
investigations that support sign-reversal superconductivity ($S_{+-}$) $^{4, 5}$ have predicted that the
order parameters of both incommensurate SDW and
SC states are compatible, because only some parts of Fermi surfaces contribute to superconductivity $^1$.

Homogeneous coexistence of the AF and SC states has been
suggested for Ba(Fe$_{1-x}$Co$_x$)$_2$As$_2$ ( Ba122 ) $^{6, 7}$ and
CaFe$_{1-x}$Co$_{x}$AsF ( Ca1111 ) $^8$. The Ca1111 series is an oxygen-free 1111 compound $^{9-11}$ that has cylindrical Fermi surfaces $^{12}$ similar to those of LaFeAsO$_{1-x}$F$_{x}$ ( La1111 ), and it exhibits an intermediate electronic phase diagram between those of the Ba122 $^{13-15}$ and La1111 $^{16, 17}$ series. An overlap between the AF and SC domes decreases with increasing interlayer distance of FeAs planes; therefore, the Ba122 series has a large overlap, the Ca1111 series has a smaller one, and the La1111 series has no overlap. For the Ba122 series, the phase overlap is large because the optimal doping level for superconductivity is located at the phase boundary between the AF and SC phases. However, for the Ca1111 series, the phase overlap is smaller than that of the Ba122 series  because the optimal doping level is away from the phase boundary $^8$. In addition, for the La1111 series, the SC phase is adjacent to the AF phase on the electronic phase diagram and thus there is no phase overlap $^{16, 17}$. Recent nuclear magnetic resonance (NMR) measurements for the La1111 series suggest that the AF and SC phases are segregated on the electronic phase diagram $^{18}$.  Although the coexistence of these phases has been suggested in the Ca1111 series, it is not certain whether the AF ordering in the crossover regime is of the same nature as that of the underdoped regime. To address this problem, it is important to investigate how a SDW state evolves upon electron doping. Therefore, we performed $^{75}$As (I=7/2) and $^{59}$Co (I=5/2) NMR measurements to investigate the evolution of the SDW and SC states upon doping in the Ca1111 series.

\begin{figure}
\includegraphics{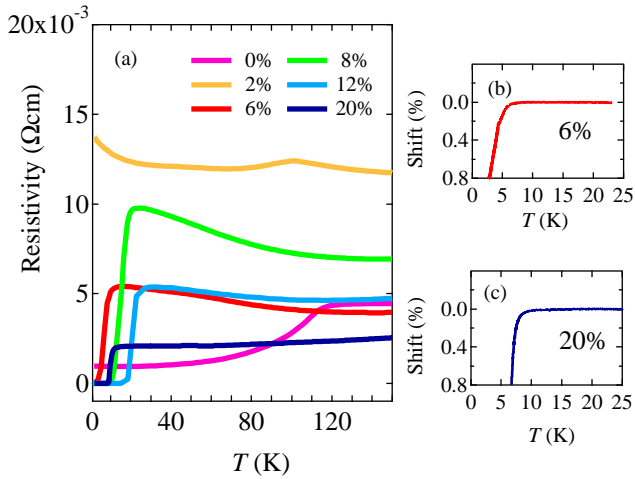}
\caption{\label{fig:epsart} (Color online) (a) Temperature dependence of the resistivity for CaFe$_{1-x}$Co$_x$AsF ($x$ = 0, 0.02, 0.06, 0.08, 0.12, and 0.20 ). (b) and (c) Detuning of an NMR tank circuit measured at 45 MHz. Bending points represent superconducting transition temperature $T_c$.   }
\end{figure}

The temperature ($T$) dependence of the resistivity is shown in Fig. 1(a). Note that the tetragonal to orthogonal phase transition manifests as an anomaly for $x \leq 0.02$, and superconductivity appears for  $x \geq 0.06$. In addition, the maximum $T_c$ is 23 K at $x = 0.12$. The SC dome determined from these resistivity measurements is shown in the electronic phase diagram ( Fig. 5) and will be discussed later. Moreover, the phase boundary between the AF and SC phases is around $x =$ 0.06 $-$ 0.07. Figures 1(b) and (c) show detuning of an NMR tank circuit for $x =$ 0.06 and 0.20; note that superconductivity is robust even for $x =$ 0.20.

Figure 2 shows $^{59}$Co NMR spectra of CaFe$_{1-x}$Co$_{x}$AsF  around the AF-SC phase boundary measured at 45.1 MHz. Homogeneous coexistence of the AF and SC phases was confirmed from the measurements for $x = 0.06$ $^8$. In general, $^{59}$Co spectra should have four edges around the central peak corresponding to I= $-\frac{1}{2} \Leftrightarrow \frac{1}{2}$, however, only
a single peak was observed in the paramagnetic ( PM ) and SC states because of some distribution of the electric field gradient ( EFG ). As shown in Fig. 2, a broad cusp-type powder pattern was observed at low temperatures for $x = 0.06$ instead of a rectangle-type powder pattern, implying that a spatially modulated spin configuration or an incommensurate SDW ( IC-SDW ) state is formed instead of a commensurate SDW state. The internal field that Co nuclei experience because of the IC-SDW ordering is expressed as $\pm h^{Co}sin(Qr)$, where $Q = \pi/n$ and 2$n$ represent the wave vector and lattice length of the spin modulation, respectively. The values  $n$ = 50 $-$ 100 and $h^{Co}$ = 6 kOe are estimated by simulating the powder pattern $^8$.  For $x \leq 0.02$, the cusp-type spin configuration was not confirmed  because a large internal field owning to some spin ordering wipes out the $^{59}$Co signals. Furthermore, the absence of the cusp-type pattern clearly indicates that the IC-SDW ordering no longer exists, because it results in $^{59}$Co signals which are free from the internal field and observable at the Larmor frequency. For $ x \geq 0.08 $, the line width remains narrow even at 4.2 K, indicating the absence of AF ordering. These results indicate that homogeneous coexistence occurs only in a narrow doping region.

\begin{figure}
\includegraphics{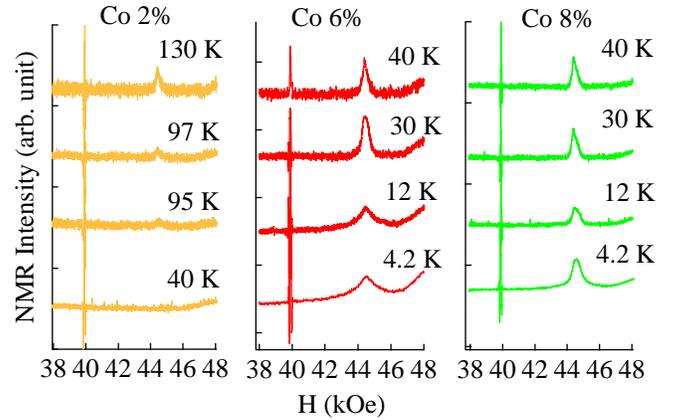}
\caption{\label{fig:epsart} (Color online) $^{59}$Co NMR spectra of CaFe$_{1-x}$Co$_{x}$AsF for $ x = $0.02, 0.06, and 0.08 measured at 45.1 MHz. The sharp signal at 40 kOe is the $^{63}$Cu signal coming from an NMR coil. The ground states for $ x = $0.02 and 0.08 are the antiferromagnetic (AF) and superconducting (SC) states, respectively. A broad cusp-type powder pattern for $ x = $0.06 shows that the spin configuration is spatially modulated.  }
\end{figure}

\begin{figure}
\includegraphics{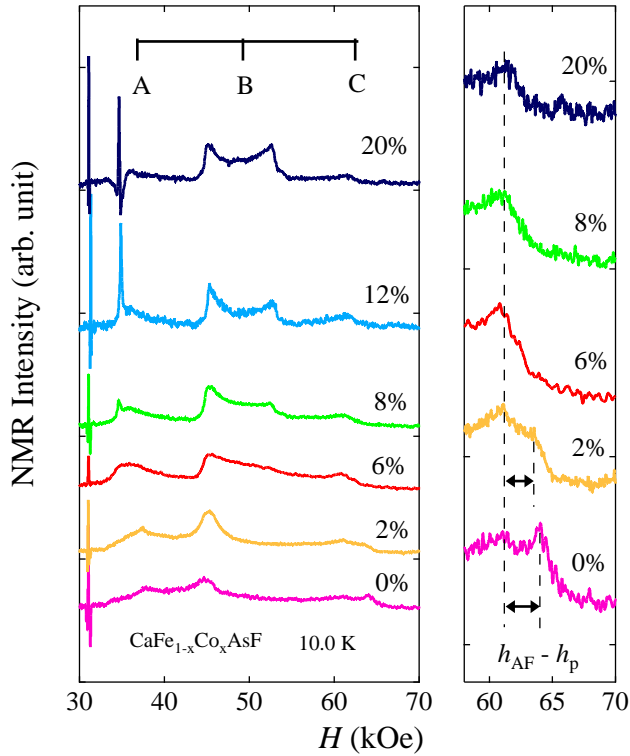}
\caption{\label{fig:epsart} (Color Online) Field-swept $^{75}$As-NMR spectra of CaFe$_{1-x}$Co$_{x}$AsF measured at 35.1 MHz and 10.0 K. The sharp signals
at 31 kOe originate from $^{63}$Cu in an NMR coil. The
sharp signals at 35 kOe for $x \geq 0.08$ originate from $^{59}$Co in
the Fe basal planes. The other signals originate from $^{75}$As. Edges A, B, and C
represent transitions I = -3/2 $\Leftrightarrow$ -1/2, I = -1/2 $\Leftrightarrow$ 1/2, and I = 1/2 $\Leftrightarrow$ 3/2, respectively. The ground state for $x \leq $0.06 is the spin ordered state. Note that spin moments suppress the higher-field edge in I = -1/2 $\Leftrightarrow$ 1/2. In addition, ordered moments manifest as edge shifts, as shown by the arrows in the right panel. The span $h_p - h_{AF}$ is given by Eqs. (4) and (5).}
\end{figure}

The evolution of the AF ordering upon electron doping is observable from $^{75}$As NMR measurements. Figure 3 shows field-swept $^{75}$As NMR spectra of CaFe$_{1-x}$Co$_{x}$AsF at 35.1 MHz and 10.0 K. For $ x \geq 0.08 $, a typical powder pattern for a PM state was observed, and sharp $^{59}$Co signals were observed at 35.1 kOe. Edges A, B and C
represent transitions I = $-3/2 \Leftrightarrow -1/2$, I = $-1/2 \Leftrightarrow 1/2$, and I = $1/2 \Leftrightarrow 3/2$, respectively. Two edges appear for I = $-1/2 \Leftrightarrow 1/2$ because of a large second-order quadrupole effect similar to those observed in other iron-based pnictides. The higher-field edge disappears for $x \leq 0.06$, because the internal field arising from AF moments wipes out the signal. For the powder samples contributing to the lower-field edge, the internal field that As nuclei experience is perpendicular to the applied field; thus, As nuclei are not affected by the internal field. Therefore, the lower-field edge remains unchanged even in the AF state. Note that, throughout the series of the spectra, we hardly found phase segregation ranging in a wide doping regime as observed from muon spin rotation ($\mu$-SR) measurements $^{19}$.

Other evidence of AF moments in CaFe$_{1-x}$Co$_{x}$AsF is the signals in the range 61 $-$ 65 kOe, as shown by the arrows in the right panel of Fig. 3. When the direction of the applied field is expressed by the zenith angle $\theta$ in spherical polar coordinates where the $z$ axis ( the maximum principle axis of EFG at $^{75}$As ) is perpendicular to FeAs planes, the spectral intensity is given as
\begin {eqnarray}
   I^{As}(h) &\propto& \int \delta ( h - (2\pi / \gamma_N)\nu(\theta)) d(cos \theta
  ), \\
  &\propto& \int  \delta (h - \frac{2\pi}{\gamma_N} \nu) |\frac{sin
\theta}{\frac{\partial \nu}{\partial\theta}}|d\nu,
\end {eqnarray} where $  h \equiv  H - (2\pi/ \gamma_N )\nu_0 $ ($\nu_0$ = 35.1 MHz) and the gyromagnetic ratio ($\gamma_N$) of $^{75}$As is 7.292 MHz/10kOe.  When converting frequency to magnetic field, one needs to multiply by the factor $\frac{2\pi}{\gamma_N}$. When EFG anisotropy is absent, the PM signals corresponding to I = $ \frac{3}{2}\Leftrightarrow \frac{1}{2}$ are given as
\begin {equation}
   \nu(\theta)= - \frac{\nu_Q}{2}(3\cos^2\theta-1),
\end {equation} where $\nu_Q = 22.6$ MHz.  Edges appear when $(\partial \nu/\partial\theta) = 0$, (i.e., $\theta = 90^{\circ}$ ) is satisfied; thus, the field position is given as $h_p = (\pi/\gamma_N) \nu_Q$. This condition is unchanged even if the EFG anisotropy is nonzero. The edge at 61 kOe corresponds to $\theta = 90^{\circ}$, as shown in the right panel of Fig. 3. The samples that have FeAs planes parallel to the applied field contribute to the peak. In the case of a uniform AF state, the internal field $\pm\Delta H \cos(\theta)$ is added to Eqs. (1) and (2). The spectral intensity $I (h)$ is given as $ I(h) \propto |3\nu_Q\cos(\theta)-\Delta\nu|^{-1} $,  where $\Delta \nu \equiv (\gamma_N/2 \pi)\Delta H$. The edge appears at $\theta = \cos^{-1} (\Delta\nu/3\nu_Q)$, namely,
\begin {eqnarray}
   h_{AF} = h_p + \Delta h_{AF}, \\
     \Delta h_{AF} = \frac{2\pi}{\gamma_N} \frac{\Delta \nu^2}{6\nu_Q}.
\end {eqnarray}
The edge is accompanied by an anomaly at the lower field $ h_p = (\pi/ \gamma_N) \nu_Q $, namely, at the edge position in a PM state. The spectral intensity at $h_p$ is $ I \propto | \Delta\nu|^{-1}$. The span between $h_p$ and $h_{AF}$ is shown by the arrows in the right panel of Fig. 3. When $\Delta\nu$ is small, $h_{AF}$ approaches or overlaps $h_p$. The value of $h_{AF}$ = 3.24 kOe results in an internal field $ \Delta H $ =24.5 kOe for $x = 0$, and $h_{AF}$ = 2.35 kOe results in $ \Delta H $ = 20.9 kOe for $x = 0.02$. The magnitude of the ordered moment $<S>$ is given as $\Delta H = g <S> A_{hf}$, where  $A_{hf}$ is the hyperfine field. Given that $A_{hf}$ $\sim$ 26 kOe/$\mu_B$ $^{20}$ and $g = 2$, $<S>$ is estimated to be 0.47$\mu_B$ and 0.40$\mu_B$ for $x = 0$ and 0.02, respectively.
 For $x = 0$, a spatially uniform spin configuration results in a sharp edge at 64 kOe, implying a commensurate spin configuration. In addition, the amplitude for $x = 0$ estimated from the neutron scattering measurements is 0.48$\mu_B$ $^{21}$, which agrees well with the present results. For $x = 0.02$, the commensurate spin configuration is maintained, although the magnitude of the spin moments is 80\% of that for $x = 0$. For $x = 0.06$, spatial spin modulation results in the spectrum that no longer exhibits the sharp edge or $\Delta h_{AF} $ given by Eqs. (4) and (5). Therefore, an incommensurate spin configuration is realized at a limited doping level, which is consistent with theoretical investigations $^{1, 2}$.


\begin{figure}
\includegraphics{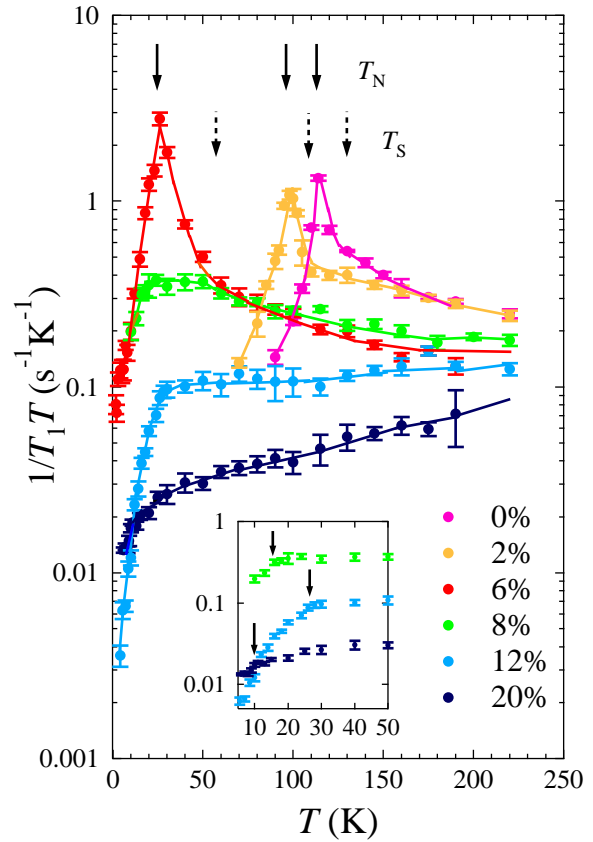}
\caption{\label{fig:wide} (Color online) Temperature dependence of $1/T_{1}T$ measured at the lower-field edge at 45 kOe in the main panel of Fig. 3. The arrows at peak positions and bending points represent antiferromagnetic ($T_N$) and structural ($T_S$) transition temperatures, respectively. The inset shows the expansion in the low-temperature range; the arrows indicate $T_c$.}
\end{figure}

The relaxation time $T_1$ was measured using the conventional
saturation-recovery method. The $T$ dependence of  $1/T_1T$ was measured at the lower-field edge ( $\sim$ 45 kOe) in the left panel of Fig. 3. Note that the bump or edge position is almost unchanged even for the AF phase. Therefore, $1/T_1T$ is successively measured even in the AF phase with the same conditions as those for the PM phase. The procedure for obtaining $1/T_1T$ was described in Ref. $^8$, and the $T$ dependence of $1/T_{1}T$ is shown in Fig. 4. The AF transition temperature $T_N$ is confirmed as a peak of $1/T_1T$ for $x \leq 0.06$ (top arrows in Fig. 4). The onset temperature of an upturn in $1/T_1T$ agrees well with the structural transition temperature $T_s$ determined from neutron scattering measurements $^{21}$. For $x = 0.08$, an upturn in $1/T_1T$ is absent, and when compared with $1/T_1T$ for $x = 0.06$, the development of the AF fluctuation is extremely suppressed and $1/T_1T$ retains the same value down to $T_c$. Therefore, for $x = 0.08$, the system undergoes a superconducting state without passing through the AF transition. This phenomenon is unique to the Ca1111 series. The maximum $T_c$ determined from $1/T_1T$ is 25 K at $ x = 0.12$, which is consistent with the resistivity measurements shown in Fig. 1. The $T$ dependence of $1/T_1T$ is proportional to $\sim T^2$ below $T_c$. For $x = 0.2$, $T_c$ is clearly observable from the detuning of an NMR tank ciurcuit as shown in Fig. 1(c); however, it is not so clear from the $T$ dependence of $1/T_{1}T$.

The relationship between quantum criticality and superconductivity is of broad interest in iron-based pnictides as well as other strongly correlated electron systems. In the Ba122 series, the quantum critical point (QCP) seems to be located at the optimal doping level $^{22}$, implying that spin fluctuation is important for the superconducting mechanism. Meanwhile, $T_s$ is very close to $T_N$ $^{13-15}$ and ultrasound absorption measurements have shown an anomaly in the elastic coefficient at $T_c$ $^{23}$, which suggests the importance of a phonon-electron coupling. It may be difficult to determine which of these mechanisms is really related to superconductivity only from the Ba122 series.
Fortunately, the structural phase boundary of the Ca1111 series is more than 20 K higher than the AF boundary at most doping levels as seen in Fig. 4, which allows one to investigate whether the QCP is really related to superconductivity. Figure 5 shows the electronic phase diagram of the Ca1111 series. This figure illustrates that the AF phase vanishes at a low doping level (the QCP $\sim$ 0.08 ) prior to the emergence of superconductivity. Note that similar  AF and SC domes have been observed in CaFe$_{1-x}$Rh$_{x}$AsF $^{24}$; therefore, the phase diagram shown in Fig. 5 retains a universal feature common to different types of dopant ions. At the QCP, as shown in Fig. 4, $1/T_1T$ hardly exhibits strong AF spin fluctuation, namely, Curie-Weiss behavior of $1/T_1T $ expected from self-consistent renormalization (SCR) theory for a two-dimensional (2D) AF system ($1/T_1T \propto \chi_Q$) $^{25}$.  Meanwhile, the structural transition survives upon considerable doping and the structural phase boundary crosses zero near the optimal doping level. These features are consistent with orbital fluctuation theory $^{26}$. As far as the Ba122 and Ca1111 series are concerned, the doping level at which the structural phase boundary crosses zero is more closely related to the optimal doping level than the QCP, although homogeneous coexistence predicted for $S_{+-}$-wave symmetry is likely to occur when the AF and SC phases overlap.

\begin{figure}
\includegraphics{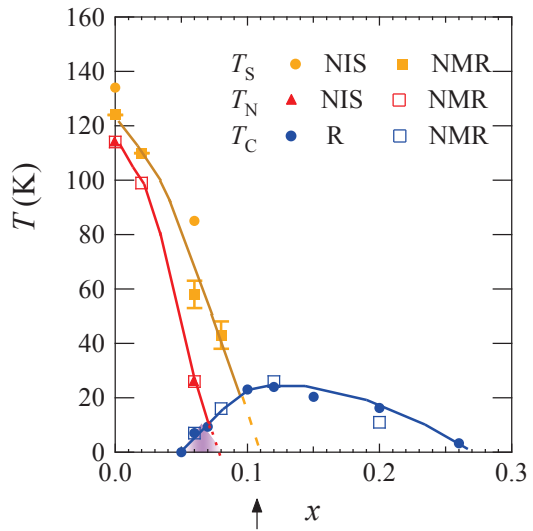}
\caption{\label{fig:epsart} (Color online) Electronic phase diagram of the Ca1111 series. The order parameters of AF and SDW states are compatible only in the small shaded region. Extrapolation of the AF boundary to zero agrees with the optimal doping level. }
\end{figure}

 In summary, we measured NMR spectra and
$1/T_1T$ for CaFe$_{1-x}$Co$_{x}$AsF, which has intermediate electronic
 and magnetic properties between LaFeAsO$_{1-x}$F$_x$ and
 Ba(Fe$_{1-x}$Co$_x$)$_2$As$_2$. We demonstrated that a uniform IC-SDW + SC state, namely a spatially homogeneous state having both the IC-SDW and SC order parameters, occurs only in a limited doping regime near the phase boundary, which suggests $S_{+-}$-wave symmetry via spin fluctuation. However, the phase diagram is not completely explained by $S_{+-}$-wave symmetry: The optimal doping level does not agree with the QCP, but it agrees with the point at which the structural phase boundary crosses zero, which is expected from $S_{++}$-wave symmetry via orbital fluctuation. These transitional features upon doping imply that both spin and orbital fluctuations are involved in the formation of superconductivity.

The NMR work is supported by a Grant-in-Aid (Grant No. KAKENHI 23340101) from the Ministry of Education, Science, and Culture, Japan. This work was supported in part by the JPSJ First Program. The authors would like to thank T. Nakano for experimental support.







\end{document}